\documentstyle[preprint,eqsecnum,aps]{revtex}

\begin{document}
\draft
\preprint{}
\title{Pseudogap and incommensurate magnetic fluctuations in YBa$_2$Cu$_3$O$_{6.6}$\\}
\author{Pengcheng Dai,$^{1}$ H. A. Mook,$^1$ and F. Do$\rm\breve{g}$an$^2$}
\address{
$^1$Oak Ridge National Laboratory, Oak Ridge, 
Tennessee 37831-6393\\
$^2$Department of Materials Science and Engineering\\ 
University of Washington, Seattle, Washington 98195\\}
\date{\today}
\maketitle
\begin{abstract}
Unpolarized inelastic neutron scattering is used to study the temperature and
wave vector dependence of the dynamical 
magnetic susceptibility, $\chi^{\prime\prime}({\bf q},\omega)$, of a well
characterized single crystal  YBa$_2$Cu$_3$O$_{6.6}$ ($T_c=62.7$ K).  
We find that  
a pseudogap opens in the spin fluctuation spectrum at temperatures well 
above $T_c$. We speculate that the appearance of 
the low frequency incommensurate fluctuations is associated with the opening of 
the pseudogap. To within the error of the measurements, a gap in the 
spin fluctuation spectrum is found in the superconducting state. 
\end{abstract}
\pacs{Keywords: High-$T_c$ superconductivity, magnetic excitations\\
\\
Pengcheng Dai\\
Oak Ridge National Laboratory\\
Oak Ridge, TN 37831-6393, USA\\
e-mail: piq@ornl.gov}

\narrowtext
It is generally recognized that the determination of the  
 spin dynamical properties of the cuprates
is crucial to the understanding of the physical origin of 
high-temperature ($T_c$) superconductivity. Neutron scattering is a unique technique
that can provide direct information on the wave vector and energy dependence of the 
imaginary part of the dynamical susceptibility, $\chi^{\prime\prime}({\bf q},\omega)$. 
Over the past several years,
 intensive experimental work on the single layer La$_{2-x}$Sr$_x$CuO$_4$ (214) 
\cite{cheong,yamada} and the 
bilayer YBa$_2$Cu$_3$O$_{7-x}$ [(123)O$_{7-x}$] 
\cite{mignod,tranquada,mook,fong,bourges,dai1,fong1} cuprates has yielded 
valuable information concerning the magnetic response in the normal and superconducting
states. To summarize the advances of the field would be difficult in this short paper, 
so we limit ourselves to recent low frequency inelastic 
neutron scattering measurements on the underdoped bilayer cuprate (123)O$_{6.6}$ at the 
Oak Ridge National Laboratory (ORNL).

The unpolarized experiments were performed  
 at the High-Flux Isotope Reactor at ORNL using the HB-3 triple-axis 
spectrometer with pyrolytic graphite (PG) as monochromator and analyzer. 
For the experiment,
we index reciprocal ($Q$) space by using the orthorhombic unit cell so that 
momentum transfers ($q_x,q_y,q_z$) in units of 
\AA$^{-1}$ are at positions 
$(H,K,L)=(q_xa/2\pi,q_yb/2\pi,q_zc/2\pi)$ reciprocal lattice units (r.l.u.). 
The sample used in present experiment 
is a well characterized (123)O$_{6.6}$ showing the onset $T_c$ of 62.7 K 
with a transition width that is 3.3 K wide \cite{dai1}.

In previous work \cite{dai1,dai2}, we have demonstrated that the magnetic response
in (123)O$_{6.6}$ is complex with incommensurate fluctuations for energies below
the commensurate resonance at low temperatures. The low frequency spin fluctuations 
change from commensurate to incommensurate on cooling with the incommensuration 
first appearing at temperatures somewhat above $T_c$. 
The incommensurate fluctuations were first discovered with the filter integration
technique \cite{dai2,mook2}. 
For a linear response function 
[$\chi^{\prime\prime}(q_x,q_y,\omega) \propto \omega F(q_x,q_y)$],
the integration range is mostly between 10 and 30 meV with the weight 
centered around 23 meV for an incident neutron energy of $\sim59$ meV. 
Thus, the observed incommensurate fluctuations with this technique should 
 mostly stem from fluctuations with an energy transfer of $\sim$23 meV.
In the subsequent triple-axis \cite{dai2} and time-of-flight measurements \cite{hayden},  magnetic
fluctuations are confirmed to be incommensurate for energies around 25 meV in the 
low temperature superconducting state. 
The intensity of the incommensurate
peaks increases on cooling below $T_c$, accompanied by a suppression of fluctuations
at the commensurate positions. However,
the energy transfers of these inelastic neutron scattering measurements 
are still large compared to that probed by the nuclear 
magnetic resonance (NMR) experiments \cite{takigawa}. 
In order to follow the incommensurate fluctuations to lower frequencies and to
compare the results with the NMR experiments, we have performed 
additional triple-axis measurements in the $(H,H,L)$ zone at energies 
lower than previously investigated \cite{dai1,dai2}.

Figure 1 shows the constant-energy scans along the $(H,H,1.8)$ direction.
The measurements were done using a collimation of 
48$^{\prime\prime}$-40$^{\prime\prime}$-40$^{\prime\prime}$-120$^{\prime\prime}$ in the usual notation.
For an energy transfer of 6 meV [see Fig. 1(a)], there is no detectable difference in the 
normal and the superconducting states. The scattering is featureless with nonobservable 
magnetic intensity around $H=0.5$ rlu. These data suggest the presence of a normal 
state spin gap (or pseudogap), consistent with previous observations \cite{mignod,tranquada}.
The same scan at an energy transfer of 10 meV is shown
 in Fig. 1(b). In contrast to the 6 meV data, the scattering in the normal 
state appears to be enhanced around the expected incommensurate positions indicated by the 
arrows. However, the weakness of the signal and the statistics of the 
data do not allow a conclusive identification about the nature of the scattering. 

Figure 2 summarizes the $(H,H,1.8)$ scans for an energy transfer of 16 meV at
various temperatures. Although the statistics of the data could still be improved, the 
temperature evolution
of the magnetic scattering is clear. At 11 K in the superconducting state 
[see Fig. 2(a)], there is 
no detectable magnetic signal around $H=0.5$, similar to the data of Fig. 1. However,
 the same scan at 75 K [see Fig. 2(b)] displays a pair of incommensurate peaks 
at $(0.5\pm\delta,0.5\pm\delta)$  ($\delta=0.054\pm0.004$) rlu. 
The observed incommensurate wave vectors are consistent with   
the value obtained using the filter integration technique \cite{dai2}. 
Warming the sample
to 125 K, the incommensuration disappears and the profile is replaced 
by a single Gaussian peak as shown in the solid line of Fig. 2(c). On further warming, 
the multi-phonon background increases significantly. 
Figure 2(d) shows
the data at 220 K, the scattering is dominated by phonons and
 the magnetic signal around $H=0.5$ has reduced drastically.

For underdoped (123)O$_{7-x}$, NMR  
experiments \cite{takigawa} suggested the presence of a pseudogap. 
The pseudogap temperature $T^\ast$ obtained from NMR 
measurements \cite{takigawa} is about 120 K for materials with a transition
temperature of $\sim60$ K. In previous neutron scattering
experiments \cite{mignod,tranquada}, the characteristic temperature 
$T^\ast$  has been associated with the opening of 
a gap (or pseudogap) in the spin fluctuations spectrum. 
The open circles in Figure 3 show the sum of the scattered intensity above the multi-phonon 
background at different temperatures obtained from constant-energy
scans as those shown in Fig. 2. The closed circles are the corresponding
susceptibility in arbitrary units after taking into account the Bose population factor.
The dynamical susceptibility peaks at $\sim$120 K, consistent with the opening of a
 pseudogap. The wave vector dependence 
of the fluctuations [see Figs. 2(c) and (d)] changes 
from commensurate above the $T^\ast$ to incommensurate below it. 
Therefore, it is  
tempting to associate the appearance of
the incommensurate fluctuations to 
the opening of a spin pseudogap. However, we stress that the measurements presented here
are still preliminary and more experiments are desirable. In particular, 
we would like to determine more precisely the temperature at which the incommensurate peaks
first appear.

In summary, we have performed preliminary measurements on the wave vector 
dependence of the low frequency spin
fluctuations for (123)O$_{6.6}$.  Our data extend the earlier measurements \cite{dai1,dai2} 
to lower 
frequencies. Two tentative conclusions are drawn from the present 
measurements: (1) There appears to be a gap in the
spin fluctuations spectrum in the low temperature superconducting state. The magnitude of
the gap is between 16 and 24 meV. (2) The appearance of
the magnetic incommensurate fluctuations in (123)O$_{6.6}$
 may be associated
with opening of a pseudogap in the spin fluctuations spectrum.

We thank G. Aeppli, V. J. Emery, S. M. Hayden, K. Levin, and D. Pines for helpful discussions. 
We have also benefited from fruitful interactions with J. A. Fernandez-Baca, R. M. Moon,
S. E. Nagler, and D. A. Tennant.
This research was supported by the US DOE under 
Contract No. DE-AC05-96OR22464 with Lockheed Martin 
Energy Research Corp.

\begin{figure}
\caption{(a) Triple-axis scans along $(H,H,1.8)$ at 6 meV in the 
normal ( $T=75$ K, $\bullet$) and superconducting ($T=11$ K, $\circ$) states.
The data were collected with a final neutron energy of 14.78 meV, and a PG filter
was placed before the analyzer. (b) Identical scans at 10 meV in two different 
temperatures. In this case, data were taken with a final neutron energy of 
30.5 meV. Arrows indicate the expected positions for incommensurate magnetic
fluctuations.
}
\label{autonum}
\end{figure}

\begin{figure}
\caption{Triple-axis scans along $(H,H,1.8)$ at 
16 meV with a final neutron energy of 35 meV for (a) 11 K,  (b) 75 K,
(c) 125 K, and (d) 220 K.  
The positions of incommensurate peaks are 
indicated by the arrows. The solid line in (b) is two Gaussians on a 
linear background.
Solid lines in (c) and(d) are Gaussian  peaks
on a linear background. 
}
\end{figure}

\begin{figure}
\caption{Temperature dependence of the magnetic scattering for an energy transfer 
of 16 meV. Since there is no observable peak at 11 K, the dynamical susceptibility is 
assumed to be zero at that temperature. For other temperatures, open circles represent
the integrated magnetic intensity above the multi-phonon background. The closed 
circles are the corresponding $\chi^{\prime\prime}(\omega)$ in arbitrary units. The
vertical dashed line is the expected pseudogap temperature $T^\ast$ and the superconducting
transition temperature is marked by the vertical arrow.}
\end{figure}

\end{document}